\documentclass[pre,showpacs,amsmath,amssymb]{revtex4}

\usepackage{graphicx}
\usepackage{dcolumn}
\usepackage{bm}

\begin{document}


\title{Correlated Binomial Models  and Correlation Structures}

\author{Masato Hisakado}
\email{masato_hisakado@standardpoors.com}
\affiliation{Standard and Poor's,
Marunouchi 1-6-5, Chiyoda-ku, Tokyo 100-0005, Japan} 
\author{Kenji Kitsukawa}
\email{kj198276@sfc.keio.ac.jp}
\affiliation{
Graduate School of Media and Governance, Keio University \\Endo 5322, Fujisawa
, Kanagawa 252-8520, Japan}

\author{Shintaro Mori}
\email{mori@sci.kitasato-u.ac.jp}
\affiliation{
Department of Physics, School of Science, Kitasato University 
\\Kitasato 1-15-1, Sagamihara, Kanagawa 228-8555, Japan
}

\date{\today}

\begin{abstract}
We discuss a general method to construct correlated binomial
distributions by imposing several consistent relations on the joint
 probability function. We obtain self-consistency relations for the
conditional correlations and conditional probabilities.
The beta-binomial distribution is derived 
by a strong symmetric assumption on the conditional correlations.
 Our derivation clarifies the 'correlation' structure  of the
 beta-binomial distribution. It is also possible to study the correlation 
structures of other probability distributions of exchangeable (homogeneous)
correlated Bernoulli random variables. We study some distribution
functions and discuss their behaviors in terms of their correlation 
structures.
\end{abstract}

\maketitle

\section{Introduction}

Incorporation of correlation $\rho$ into Bernoulli random variables
$X_{i} (i=1,2,\cdots N)$ taking the value 1 with probability $p$ and 
taking the value 0 with probability $1-p$  has long history and have 
been widely discussed  in a variety of areas of science, mathematics and 
engineering.  Writing the expectation value of a random variable 
$A$ as $<A>$, the correlation $\rho$ between $X_{i}$ and $X_{j}$
is defined as 
\begin{equation}
\rho=\mbox{Corr}(X_{i},X_{j}) 
=\frac{<X_{i}X_{j}>-<X_{i}><X_{j}>}{\sqrt{<X_{i}>(1-<X_{i}>)<X_{j}>
(1-<X_{j}>)}}  \label{pearson}. 
\end{equation}
If there are no correlation between the random variables,
 the number $n$ of the variables taking the value 1 obeys the binomial 
probability distribution b$(N,p)$. The necessity of the correlation 
$\rho$ comes from the facts that there are many phenomena where 
dependency structures in the random events are crucial
or are necessary for the explanation of experimental data.

For example, in biometrics, the teratogenic or toxicological
effect of certain compounds was studied
\cite{Griffiths,Williams,Kupper}. 
The interest resides in the number of affected fetuses or 
implantation in a litter.
One parameter models, such as the Poisson distribution and 
binomial distributions provided poor fits to the experimental data.
A two-parameter alternative to the above distributions, beta-binomial
distribution (BBD), has been proposed \cite{Griffiths,Williams}.
 In the model, the probability $p'$ of the binomial distribution 
 b$(N,p')$ 
is also a random variable and obeys the beta distribution 
Be$(\alpha,\beta)$.
\begin{equation}
 P(p')=\frac{p^{'\alpha-1}(1-p')^{\beta-1}}{\mbox{B}(\alpha,\beta)}.
\end{equation}
The resulting distribution has probability function 
\begin{equation}
P(n)={}_{N}C_{n} \cdot
 \frac{\mbox{B}(\alpha+n,N+\beta-n)}{\mbox{B}(\alpha,\beta)}   \label{beta_B}. 
\end{equation}
The mean $\mu$ and variance $\sigma^{2}$ of the BBD are
\begin{equation}
 \mu=N p \hspace*{0.3cm}\mbox{and}
\hspace*{0.3cm} \sigma^{2}=N pq(1+N\theta)/
(1+\theta)
\end{equation}
where
\begin{equation}
p=\frac{\alpha}{\alpha+\beta}
\hspace*{0.3cm}\mbox{,}
\hspace*{0.3cm} q=1-p=\frac{\beta}{\alpha+\beta}
\hspace*{0.3cm}\mbox{and}
\hspace*{0.3cm}\theta=\frac{1}{\alpha+\beta}.       
\end{equation}
$\theta$ is a measure of the variation in $p'$ and is called as
``correlation level'' \cite{Bakkaloglu}. 
The case of pure binomial distribution
corresponds to $\theta=0$. However, true ``correlation'' of the BBD is
given as
\begin{equation}
\rho=\frac{1}{\alpha+\beta+1}.   \label{corr_BBD}
\end{equation}
The derivation of the relation is straightforward. 
If we denote the sum of $X_{i}$ as $S=\sum_{i=1}^{N}X_{i}$, we can write 
as  $<X_{i}X_{j}>=<S^{2}-S>/N(N-1)$ and $<X_{i}>=<X_{j}>=<S>/N$. 
From eq.(\ref{pearson}) and the
results for BBD, we obtain eq.(\ref{corr_BBD}).  We rewrite the variance
$\sigma^{2}$ as
\begin{equation}
\sigma^{2}=N pq+N(N-1)pq \cdot \rho.
\end{equation}

In the area of computer engineering, in the context of the
design of survivable storage system, the modeling of the correlated
failures  among storage nodes is a hot topic \cite{Bakkaloglu}. 
In addition to BBD, a correlated 
binomial model based on conditional failure probabilities 
has been proposed. The same kind of  correlated binomial distribution 
based on conditional
probabilities has also been introduced
in financial engineering. There, credit portfolio modeling   
has been extensively studied \cite{Schonbucher,Frey}.
In particular, the modeling  default correlation plays central
role in the pricing of portfolio credit derivatives, which 
are developed in order to manage the risk of joint default or the 
clustering of default. As a default distribution 
model for homogeneous (exchangeable) credit portfolio where the assets' default
probabilities and default correlations are uniform and denoted as
$p$ and $\rho$, Witt has introduced a correlated
binomial model based on the conditional default 
probabilities $p_{n}$ \cite{Witt}.
 Describing the defaulted (non-defaulted) state
of i-th asset by $X_{i}=1\hspace*{0.2cm} (X=0)$ and 
the joint default probability function
by $P(x_{1},x_{2},\cdots,x_{N})$, 
 $p_{n}$ are defined as
\begin{equation}
p_{n}=<X_{n+1}| \prod_{n'=1}^{n}X_{n'}=1> \label{cond_p}.
\end{equation}
Here $<A|B>$ means the expectation value of a random variable
$A$ under the condition that $B$ is satisfied. The expectation value
of $X_{i}$ signifies the default probability and the condition 
$\prod_{n'=1}^{n}X_{n'}=1$ corresponds to the situation where
the first $n$ assets among $N$ are defaulted. $p_{0}=p$ and
from the homogeneity (exchangeability ) assumption, any $n$ assets  among $N$ 
can be chosen in the $n$ default condition $\prod_{n'=1}^{n}X_{n'}=1$. 
$X_{n+1}$ in eq.(\ref{cond_p}) is also substituted by anyone
which is not used in the $n$ default condition. 

In order to fix the joint default probability function completely, 
it is necessary to impose $N$ conditions on them from the homogeneity 
assumption. Witt and the authors
have imposed the
following condition on the conditional correlations 
\cite{Witt,Mori}.
\[
\mbox{Corr}(X_{n+1},X_{n+2}|\prod_{n'=1}^{n}X_{n'}=1)=\rho \exp(-\lambda
n)\equiv \rho_{n}. \label{cond_rho}
\]
Here Corr$(A,B|C)$ means the correlation between the random variable $A$
and $B$ under the condition $C$ is satisfied. From them, recursive 
relations for $p_{n}$ are obtained and $p_{n}$ are
calculated as 
\[
p_{n}=1-(1-p)\prod_{n'=0}^{n-1}(1-\rho_{n'}). 
\]
The joint default probability function and the default distribution 
function $P_{N}(n)$ has been expressed with these $p_{n}$ explicitly. 
However, the expression has many $\pm$ contributions and it is not an easy task
to evaluate them for $N\ge 100$. In addition, the range of parameters
$p$ and $\rho$ are also restricted and one cannot study the large
correlation regime. Furthermore, for $p=0.5$ case, the distribution does
not have the $Z_{2}$ symmetry as $P_{N}(n)=P_{N}(N-n)$. The distribution
has irregular shape and for some choice of parameters, it shows singular
rippling.

In this paper, we propose a general method to construct correlated 
binomial models (CBM) based on the consistent conditions on the conditional
probabilities and the conditional correlations. With the method, it is 
possible to study the correlation structure for any probability 
distribution function for
exchangeable correlated Bernoulli random variables. 
The organization of the paper is as follows.
In section \ref{Model}, we introduce conditional probabilities $p_{ij}$
and conditional correlations $\rho_{ij}$ and show how to construct CBMs. 
We prove  that the construction is self-consistent. In addition, in order
to assure the probability conservation or the normalization, 
 the conditional correlations
and the probabilities should satisfy self-consistent relations.
We also calculate the moments $<n^{k}>$ of the model.
In the course, we introduce a linear operator $H$ which gives the
joint probabilities in the ``binomial'' expansion of $(p+q)^{N}$.    
Section \ref{Solution} is devoted to some solutions of the  
self-consistent relations. We obtain the beta-binomial distribution 
(BBD) with strong symmetric assumptions on the conditional correlations.
For other probability distribution functions which include the Witt's
model and the distributions constructed by 
the superposition of the binomial distributions 
(Bernoulli mixture model),
 we calculate $p_{ij}$ 
and $\rho_{ij}$. We study the probability
distribution functions for these solutions from the viewpoint of
their correlation structures $\rho_{ij}$.  
We conclude with some remarks and
future problems in section \ref{Conclusion}.

\section{Correlated Binomial Models and Their Constructions}
\label{Model}
 
\begin{figure}[htbp]
\begin{center}
\includegraphics[width=9cm]{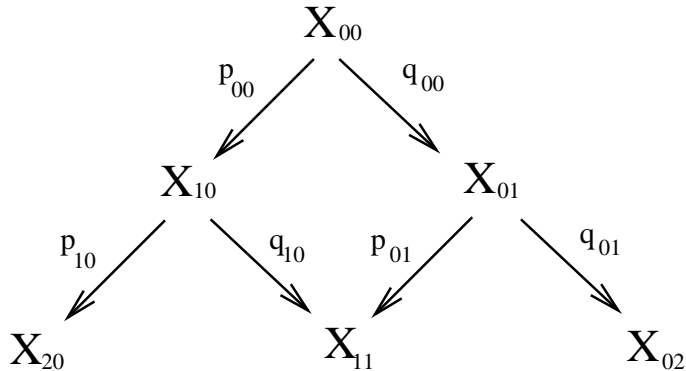}
\end{center}
\caption{Pascal's triangle like representation 
of $X_{ij}$ and $p_{ij},q_{ij}$ up to $i+j\ \le 2$.
$X_{00}=<1>$, $X_{10}=<X_{1}>=p$, $X_{01}=<1-X_{1}>=1-p=q$ etc.}
\label{pascal}
\end{figure}

In this section, we construct the joint probabilities and the 
distribution functions of CBMs.
We introduce the following definitions. The first one is the products
of $X_{i}$ and $1-X_{j}$ and they include all observables of the model.
\begin{equation}
\Pi_{ij}=\prod_{i'=1}^{i}X_{i'}\prod_{j'=i+1}^{i+j}(1-X_{j'})
\end{equation}
The following definitions are their unconditional and conditional 
expectation values (see Figure 
\ref{pascal}.).
\begin{eqnarray}
X_{ij}&=&<\Pi_{ij}> \\
p_{ij}&=&<X_{i+j+1}|\Pi_{ij}=1>=\frac{X_{i+1j}}{X_{ij}}
\\
q_{ij}&=&<1-X_{i+j+1}|\Pi_{ij}=1>=\frac{X_{ij+1}}{X_{ij}}
\end{eqnarray}
$X_{00}=1$,
$X_{10}=p$ and $X_{01}=1-p=q$. Furthermore, the relation 
$p_{ij}+q_{ij}=1$ should hold for any $i,j$, because of the identity
$<1|\Pi_{ij}=1>=<X_{i+j+1}+(1-X_{i+j+1})|\Pi_{ij}=1>=1$. 
All informations are contained in $X_{ij}$. The 
joint probability $P(x_{1},x_{2},\cdots,x_{N})$ with 
$\sum_{i'=1}^{N}x_{i'}=n$ is given by $X_{n N-n}$ and 
the distribution function $P_{N}(n)$ is also calculated as
\begin{equation}
P_{N}(n)={}_{N}C_{n} \cdot X_{n N-n}.
\end{equation}
In order to estimate $X_{ij}$, we need to calculate the products of $p_{kl}$ and $q_{kl}
$ from $(0,0)$ to $(i,j)$. As the path, we can choose anyone and the
product must not depend on the choice. This property is guaranteed by 
the next
condition on $p_{ij}$ and $q_{ij}$ as
\begin{equation}
q_{i+1j}\cdot p_{ij}=p_{ij+1}\cdot q_{ij}=\frac{X_{i+1 j+1}}{X_{ij}}.
 \label{consist}
\end{equation}
In order for $p_{ij}$ and $q_{ij}$ to satisfy these conditions,
we introduce the following conditional correlations
\begin{equation}
\mbox{Corr}(X_{i+j+1},X_{i+j+2}|\Pi_{ij}=1)=\rho_{ij}.
\end{equation}
We set $\rho_{00}=\rho$.
$(1-X_{i+j+1})$ and $(1-X_{i+j+2})$ are also correlated with the same 
strength and the following relations hold.
\begin{equation}
\mbox{Corr}((1-X_{i+j+1}),(1-X_{i+j+2})|\Pi_{ij}=1)=\rho_{ij}.
\end{equation}
From these relation, we obtain the recursive relations for 
$p_{ij}$ and $q_{ij}$ as
\begin{eqnarray}
p_{i+1j}&=&p_{ij}+(1-p_{ij})\rho_{ij} \nonumber \\
q_{ij+1}&=&q_{ij}+(1-q_{ij})\rho_{ij}.   \label{recursive}
\end{eqnarray}
If we assume the identity $p_{ij}+q_{ij}=1$, 
we obtain
$q_{ij}=1-p_{ij}$, $q_{i+1j}=1-p_{i+1j}=(1-p_{ij})(1-\rho_{ij})$
and $p_{ij+1}=1-q_{ij+1}=p_{ij}(1-\rho_{ij})$.
Then $q_{i+1j}\cdot p_{ij}=p_{ij+1}\cdot q_{ij}
=p_{ij}(1-p_{ij})(1-\rho_{ij})$ holds and we see that 
the above consistency relation (\ref{consist}) does hold.

\begin{figure}[htbp]
\begin{center}
\includegraphics[width=8cm]{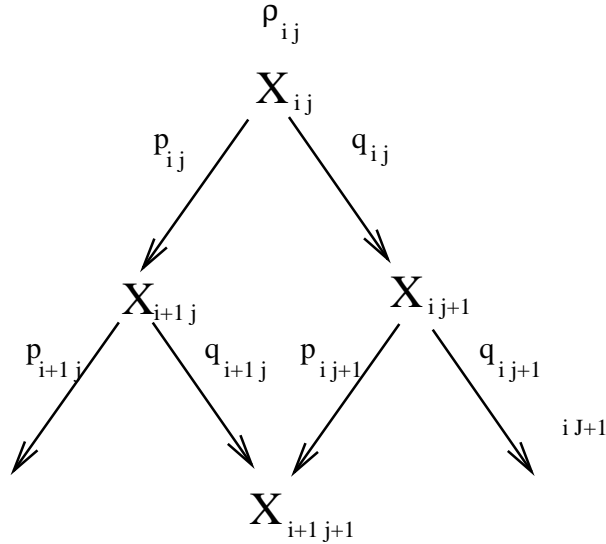}
\end{center}
\caption{Proof of the commutation relation 
$q_{i+1j}\cdot p_{ij}=p_{ij+1}\cdot q_{ij}$.
}
\label{commutation}
\end{figure}

The remaining consistency relations or the probability conservation identity
is $p_{ij}+q_{ij}=1$. We prove the identity by the inductive method.
For $i=j=0$, the identity holds trivially 
as $p_{00}+q_{00}=p+q=1$. For $j=0$ or $i=0$, $q_{i0}$ and $p_{0j}$ are 
calculated as $q_{i0}=1-p_{i0}$ and $p_{0j}=1-q_{0j}$ and the identity
also holds trivially. Then we assume $p_{ij-1}+q_{ij-1}=1$ and prove
the identity $p_{ij}+q_{ij}=1$. From the recursive equations 
(\ref{recursive}) on $p_{ij}$
and $q_{ij}$, we have the following relations.
\begin{eqnarray}
&&1=p_{ij}+q_{ij}  \nonumber \\
&&=p_{i-1j}+(1-p_{i-1 j})\rho_{i-1j}+(1-p_{ij-1})
+p_{ij-1}\rho_{ij-1} .
\end{eqnarray}
For the identity to be satisfied, the conditional correlation
$\rho_{ij-1}$ and $\rho_{i-1j}$ 
must satisfy the following relations.
\begin{equation}
p_{i-1j}-p_{ij-1}=-(1-p_{i-1j})\rho_{i-1j}-p_{ij-1}\rho_{ij-1}. 
\label{consist2}
\end{equation}
If the conditional correlations $\rho_{ij}$ are fixed so as to
satisfy the relations, the model becomes self-consistent. In other
words, it guarantees the normalization of the resulting probability
distribution.

\begin{figure}[htbp]
\begin{center}
\includegraphics[width=8cm]{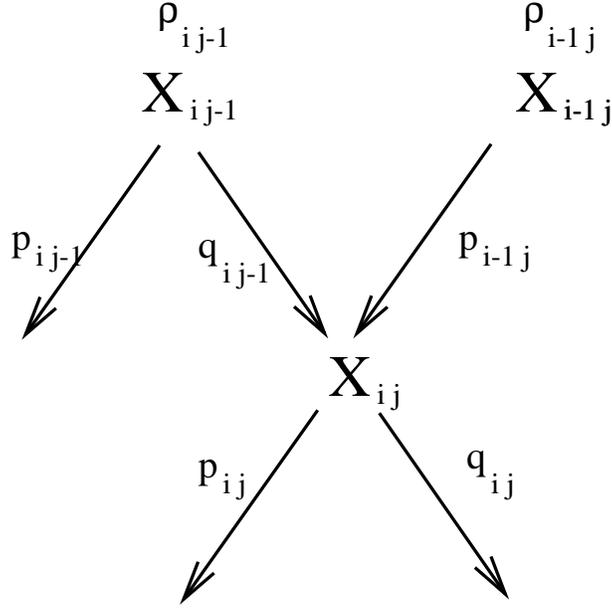}
\end{center}
\caption{Picture for the $p_{ij}+q_{ij}=1$ condition.}
\label{pq1}
\end{figure}

We estimate the moments of CBM. For the purpose, we 
introduce following operators $H$ and $D_{k}$. 
The former one is a linear operator $H$ which maps
polynomial in $p,q$ to joint probabilities $\in R$. 
By its linearity, we only need to 
fix its action on monomial 
$p^{i}q^{j}$ as
\begin{equation}
H[p^{i}q^{j}]=p_{00}p_{10}\cdots p_{i-1 0}q_{i0}q_{i1}\cdots q_{i j-1}.
\end{equation}
The joint probability $X_{n N-n}$ is expressed as $X_{n
N-n}=H[p^{n}q^{N-n}]$. Here  we choose the far left path from $(0,0)$ to 
$(n,N-n)$ on the Pascal's triangle (See Figure \ref{pascal}).
The action of $H$ on the 
binomial expansion $(p+q)^{N}=1^{N}$ can be interpreted as the
 probability distribution and its
normalization condition.
\begin{equation}
1=H[1^{N}]=H[(p+q)^{N}]=
\sum_{n=0}^{N}{}_{N}C_{n}\cdot
H[p^{n}q^{N-n}]=
\sum_{n=0}^{N}{}_{N}C_{n}\cdot
X_{n N-n}
\end{equation}
In order to calculate the moments of CBM, it is necessary to put 
$n^{k}$ in the above summation. Instead, we will put $n(n-1)(n-2)\cdots 
(n-k+1)$ and introduce the following differential operators 
$D_{k}$.
\begin{equation}
D_{k}=\sum_{0\le i_{1},i_{2},\cdots \le N-1}
^{i_{1}\neq i_{2},i_{1}\neq i_{3},\cdots,i_{k-1}\neq i_{k}}
p_{i_{1}0}p_{i_{2}0}
\cdots
p_{i_{k}0}\frac{\partial^{k}}{\partial p_{i_{1}0}\partial p_{i_{2}0}
\cdots \partial p_{i_{k}0}}
\end{equation}
The action of $D_{k}$ on $X_{n N-n}$ for $n\ge k$ is
\begin{equation}
D_{k}X_{n N-n}=n(n-1)(n-2)\cdots (n-k+1)X_{n N-n}.
\end{equation}
On the other hand, the same expression can be obtained as
\begin{equation}
H[p^{k}\frac{d^{k}}{dp^{k}} p^{n}q^{N-n}]
=H[n(n-1)(n-2)\cdots (n-k+1) p^{n}q^{N-n}]
=n(n-1)(n-2)\cdots (n-k+1)X_{n N-n}.
\end{equation}
This relation defines the action of $D_{k}$ on the operator 
$H$ with any polynomial $f(p,q)$ as
\begin{equation}
D_{k}H[f(p,q)]=H[p^{k}\frac{d^{k}}{dp^{k}} f(p,q)]. \label{DF}
\end{equation}
The calculation of the expectation value of $n(n-1)\cdots
(n-k+1)$ is performed by the action of operator $D_{k}$ on the
binomial expansion of $H[1^{N}]=H[(p+q)^{N}]$.
\begin{equation}
D_{k}H[(p+q)^{N}]=\sum_{n=0}^{N} {}_{N}C_{n}\cdot D_{k}X_{n N-n}
\end{equation}
The right hand side is nothing but the expectation value 
$<n(n-1)(n-2)\cdots (n-k+1)>$. The left hand side is calculated by using 
eq.(\ref{DF}) as
\begin{eqnarray}
&&D_{k}H[(p+q)^{N}]=H[p^{k}\frac{d^{k}}{dp^{k}}(p+q)^{N}]
=N(N-1)(N-2)\cdots (N-k+1)H[p^{k}(p+q)^{N-k}] \nonumber \\
&=&
N(N-1)(N-2)\cdots (N-k+1)H[p^{k}]
=N(N-1)(N-2)\cdots (N-k+1)p_{00}p_{10}p_{20}\cdots p_{k-1 0}.
\end{eqnarray}
We obtain the relation,
\begin{equation}
<n(n-1)(n-2)\cdots (n-k+1)>=
N(N-1)(N-2)\cdots (N-k+1)p_{00}p_{10}p_{20}\cdots p_{k-1 0}
\end{equation}
From the relation, we can estimate the moments of CBM.

\section{Beta-binomial Distribution and Other Solutions}
\label{Solution}

In the previous section, we have derived self-consistent equations
for $p_{ij}$ and $\rho_{ij}$. They are summarized as
\begin{eqnarray}
p_{i+1j}&=&p_{ij}+(1-p_{ij})\rho_{ij} \label{consist_p}\\
p_{ij+1}&=&p_{ij}-p_{ij}\rho_{ij}   \label{consist_q} \\
p_{i-1j}-p_{ij-1}&=&-(1-p_{i-1j})\rho_{i-1j}-p_{ij-1}\rho_{ij-1}
\label{consist3}. 
\end{eqnarray}
In this section, we show several solutions to these equation.
We note, if one knows joint probabilities $X_{ij}$, from the
definitions for $p_{ij}$ and $q_{ij}$, we can estimate $p_{ij}$.
Then $\rho_{ij}$ are estimated from the recursive equation 
(\ref{consist_p}). In addition, we interpret the 
behaviors of the solutions from the viewpoint of 
correlation structures.

\subsection{Beta-binomial Distribution}

In order to solve the above relations on $\rho_{ij}$ and $p_{ij}$,
we use the symmetry viewpoint. For $p=\frac{1}{2}$ case, the 
model should have particle-hole duality between $X$ and $1-X$ or 
$Z_{2}$ symmetry. Then $\rho_{ij}=\rho_{ji}$ should hold.
We put  stronger assumption that for any $p$, the system has the $Z_{2}$
symmetry and $\rho_{ij}$ depends on $i,j$ only through the
combination $n=i+j$. With a suitable choice of indexes $i\to i+1$ and 
$j=n-i$,
eq.(\ref{consist3}) reduces to
\begin{equation}
p_{i n-i}-p_{i+1 n-i-1}=\rho_{n}(-1+p_{i n-i}-p_{i+1 n-i-1}).
\end{equation}
From this relation, we see that $p_{ij}$ with the same $n=i+j$ consist
a arithmetic sequence with the common difference $\Delta_{n}$.
\begin{equation}
p_{i+1 n-i-1}-p_{i n-i}=\Delta_{n}.
\end{equation}
$\Delta_{n}$ satisfy the following equation
\begin{equation}
\Delta_{n}=\rho_{n}(1+\Delta_{n}).
\end{equation}
$\rho_{n}$ can be solved with $\Delta_{n}$ as
\begin{equation}
\rho_{n}=\frac{\Delta_{n}}{1+\Delta_{n}}.
\end{equation}
From the relation (\ref{consist_p}) for $p_{ij}$, we obtain the
following recursive relation for
$\rho_{n}$ as,
\begin{equation}
\rho_{n}=\frac{\Delta_{n}}{1+\Delta_{n}}=\frac{\Delta_{n-1}(1-\rho_{n-1})}{
1+\Delta_{n-1}(1-\rho_{n-1})}=\frac{\rho_{n-1}}{1+\rho_{n-1}}.
\end{equation}
The explicit form for $\rho_{n}$ and $\Delta_{n}$ are
\begin{equation}
\rho_{n}=\frac{\rho}{1+n\rho} \hspace*{0.3cm} \mbox{and}
\hspace*{0.3cm} \Delta_{n}=\rho_{n-1}.
\end{equation}
Then $p_{ij}$ and $q_{ij}$ can be obtained explicitly and the results 
are
\begin{eqnarray}
p_{ij}&=&p_{i+j 0}-j\Delta_{i+j}=\frac{p(1-\rho)+i\rho}{1+(i+j-1)\rho} \\
q_{ij}&=&1-p_{ij}=\frac{q(1-\rho)+j\rho}{1+(i+j-1)\rho}.
\end{eqnarray}
$X_{n N-n}$ are then obtained  by taking the products of producing
these conditional probabilities  from $(0,0)$ to $(n,N-n)$
\begin{equation}
X_{n N-n}=\prod_{i=0}^{n-1}p_{i0} 
\prod_{j=0}^{N-n-1}q_{nj}.
\end{equation}
Putting the above results for $p_{ij}$ and $q_{ij}$ into them, we obtain
\begin{equation}
X_{n N-n}=\frac{\prod_{i=0}^{n-1}(p(1-\rho)+i\rho)
\prod_{j=0}^{N-n-1}(q(1-\rho)+j\rho)}{\prod_{k=0}^{N-1}(1+(k-1)\rho)}
\label{Xij}.
\end{equation}
Here $q=1-p$. 
By multiplying the binomial coefficients ${}_{N}C_{n}$, we obtain the 
distribution function $P_{N}(n)$ as
\begin{equation}
P_{N}(n)={}_{N}C_{n} \cdot X_{n N-n}.
\end{equation}
 This distribution is nothing but the beta-binomial
distribution function (see eq.(\ref{beta_B})) with 
suitable replacements $(p,\rho) \leftrightarrow (\alpha,\beta)$ . 

\subsection{Moody's Correlated Binomial Model}

In the original work by Witt, he assumed $\rho_{i,0}=\rho$ for all 
$i$ \cite{Witt}. We call this model as Moody's Correlated Binomial
(MCB) model.
The above consistent 
equations are difficult to solve and the available analytic expressions 
are those for $p_{i0}$ as $p_{i0}=1-(1-p)(1-\rho)^{i}$. With the result,
we only have a formal expression for $X_{ij}$ as
\begin{eqnarray}
X_{ij}&=&<\Pi_{ij}>=<\prod_{i'=1}^{i}X_{i'}\prod_{j'=i+1}^{i+j}
(1-X_{j'})> \nonumber \\
&=&\sum_{k=0}^{j}(-1)^{k}{}_{j}C_{k}<\prod_{i'=1}^{i+k}X_{i'}>
=\sum_{k=0}^{j}(-1)^{k}{}_{j}C_{k} \cdot p_{i+k 0} \label{formal}.
\end{eqnarray}
With this expression, it is possible to estimate $p_{ij}$,$q_{ij}$
 and $\rho_{ij}$ from their definitions. 
However, equation (\ref{formal}) contains 
${}_{j}C_{k}(-1)^{k}$ and as $N$ becomes large, it becomes difficult
to estimate them. With the above choice for $\rho_{i0}=\rho$, 
it is possible to set $N=30$. If $\rho_{i0}$ damps as 
$\exp(-\lambda i)$ with some positive $\lambda$, we can set at most 
$N=100$ for small values of $\rho$ and $p$.

\subsection{Mixed Binomial Models: Bernoulli Mixture Models}

Bernoulli mixture model with some mixing probability
distribution function $f(p)$, the expression for the joint probability
function $X_{ij}$ is calculated with 
\begin{equation}
X_{ij}=<\Pi_{ij}>=\int_{0}^{1} dp f(p)p^{i}(1-p)^{j}.  \label{mix}
\end{equation}
If we use the beta distribution for $f(p)$, we obtain eq.(\ref{Xij}).
However, this does not mean that 
it is trivial to solve the consistent equations
 with the assumption $\rho_{ij}=\rho_{i+j}$ and
 obtain the BBD. 
The consistent equations completely determine any correlated binomial
distribution for exchangeable Bernoulli random variables. 
Every correlated binomial distributions obey the relations.
With the assumption $\rho_{ij}=\rho_{i+j}$, we are automatically
lead to the BBD. That is, the probability distribution with the symmetry
 $\rho_{ij}=\rho_{i+j}$, we prove that it is the BBD. 
No other probability distribution has the symmetry.

Here we consider the relation between
CBM and Bernoulli Mixture model. According to De Finetti's theorem,
 the probability distribution of any infinite exchangeable
Bernoulli random variables can be  expressed by a mixture of the
binomial distribution \cite{DeFinetti}. CBM in the $N\to \infty$ limit
 should be expressed by such a mixture.  From eq.(\ref{mix}), we have the
relation $P(x_{1}=1,x_{2}=1,\cdots,x_{k}=1)=X_{k0}=\int f(p)p^{k}dp$. 
$X_{k0}$ is expressed as $X_{k0}=p_{00}p_{10}\cdots p_{k-1 0}$, we 
have a correspondence between the moments of $f(p)$ and a CBM.
That is, if one knows $p_{i0}$ for any $i$, we know the 
mixing function $f(p)$ and vice versa. This correspondence 
shows the equivalence of CBM and the Bernoulli
mixture model in the large $N$ limit. 
But CBMs with finite $N$ can describe  probability distribution more widely.
In the Bernoulli mixture model, the variance of $p$ is positive
and the correlation $\rho$ cannot be taken negative. In CBM, we can set
$\rho$ negative for small system size $N$.  In addition, CBM is
useful to
construct the probability distribution and 
discuss about the correlation structure.
Particularly  we can understand the symmetry of the solution.
For example, we want to have $Z_2$ symmetry distributions.
In the Bernoulli mixture model, we need to impose on $f(p)$ as
\begin{equation}
\int_{0}^{1}f(p)(p-0.5)^{2k+1}dp=0,
\end{equation}
where $k=1,2,\cdots$.
On the other hand, in CBM, we only need to seek a solution 
with $p_{ii}=q_{ii}=\frac{1}{2}$.
This simple constraint is useful in the construction and
in the parameter calibration of CBMs.

As other mixing functions $f(p)$, we consider the cases which correspond to 
the long-range Ising model with some strength of magnitude 
of correlation $\rho >0$.
It has some correlation only in the regime where 
the probability distribution for the magnetization $p(m)$
has two peaks at $m_{1},m_{2}$ for  $T < T_{c}$ \cite{Kitsukawa}.
If the system size $N$ is large enough, the distribution can be
approximated with the superposition of two  binomial distributions.
If we take $N \to \infty$ for $T <T_{c}$, 
the system loses its ergodicity  
and the phase space breaks up into two space with $m >0$ and $m<0$ 
\cite{Goldenfeld}  and the correlation disappears. 
Even if there appears two peaks 
in $p(m)$, only one of them represents the real equilibrium state.

The precise values of $m_{1}$ and $m_{2}$ depend on the model
 parameters, we consider the cases which correspond to
$p=0.5$ ($Z_{2}$ symmetric case) and  $p\simeq 0$.
For the $Z_{2}$ symmetric case, there is no external field and
$m_{1}=-m_{2}$ holds. Between the  Bernoulli random variable $X$ and the
Ising Spin variable $S$, there exists a mapping  
$X=\frac{1}{2}(1-S)$.
$f(p)$ has two peaks at $p$ and $q=1-p$ with the same height.
On the other hand, for $T \simeq 0$ and infinitely weak positive external
field case $\sim O(\frac{1}{N})$, $p(m)$ has one tall peak at $m_{1} \simeq 1$ 
and another short peak at $m_{2} \simeq -1$. In the language of the
Bernoulli random variable case, $f(p)$ has a tall peak at $p'=p'' \simeq 0$
and a short peak at $p' \simeq 1$. We consider the following mixing 
functions and call them Two-Binomial models.

\begin{itemize}
\item $f(p')=\frac{1}{2}\delta(p'-p)+\frac{1}{2}\delta(p'-q)$ with $q=1-p$.

This mixing function corresponds to the long-range Ising model with
      $Z_{2}$ symmetry and $\rho>0$. 
$X_{ij}$ are given as
\begin{equation}
X_{ij}=\frac{1}{2}(p^{i}q^{j}+p^{j}q^{i})
\label{basic}.
\end{equation}
$p_{ij}$ and $\rho_{ij}$ are calculated easily as
\begin{eqnarray}
p_{ij}&=&\frac{p^{i+1}q^{j}+p^{j}q^{i+1}}{p^{i}q^{j}+p^{j}q^{i}}
   \\
\rho_{ij}&=&\frac{p^{i+j}q^{i+j}(p-q)^{2}}{p^{i+j}q^{i+j}(p^{2}+q^{2})
+qp(p^{2i}q^{2j}+q^{2i}p^{2j})}.
\end{eqnarray}
This solution has the $Z_{2}$ symmetry $\rho_{ij}=\rho_{ji}$.

\item $f(p')=\frac{p^k}{p^k+q^k}\delta(p'-p)
+\frac{q^k}{p^k+q^k}\delta(p'-q)$ with $q=1-p$.

This is the modified version of the above solution with a parameter 
$k=0,1,\cdots$. 
If we set $k=0$, it is nothing but the above solution. 
$X_{ij}$ are given as
\begin{equation}
X_{ij}=\frac{1}{p^k+q^k}(p^{i}q^{j}p^{k}+p^{j}q^{i}q^{k}).
\end{equation}
$p_{ij}$ and $\rho_{ij}$ are
\begin{eqnarray}
p_{ij}&=&\frac{p^{i+k+1}q^{j}+p^{j}q^{i+k+1}}{p^{i+k}q^{j}+p^{j}q^{i+k}}
   \\
\rho_{ij}&=&\frac{p^{i+j+k}q^{i+j+k}(p-q)^{2}}{p^{i+j+k}q^{i+j+k}(p^{2}+q^{2})
+qp(p^{2i+2k}q^{2j}+q^{2i+2k}p^{2j})}.
\end{eqnarray}
If we denote $C_{1}=\frac{p^{k}}{p^{k}+q^{k}}$, $C_{2}=\frac{q^{k}}
{p^{k}+q^{k}}$, then the mixing function becomes $f(p')=C_{1}\delta(p'-p)
+C_{2}\delta(p'-q)$. This solution may look trivial.
 One  obtain this solution 
using the parallel shift of the above solution (\ref{basic}).
We replace $X_{ij}$ with $X_{i+k j}$ in eq.(\ref{basic}) and obtain the 
solution. Such a parallel shift may give birth to another solution, we 
would like to note it here.

\item $f(p')=(1-\alpha)\delta(p'-p'')+\alpha \delta(p'-1)$.

This mixing function corresponds to the long-range Ising model without 
$Z_{2}$ symmetry, $<S_{i}> \simeq 1$ and $\rho>0$.  
We call the model as Binomial plus (B+) model, because  it 
is a binomial distribution plus one small peak at $n=N$.
Between $p,\rho$ and $p'',\alpha$, we have the relations
\begin{equation}
p=\alpha+(1-\alpha)p'' \hspace*{0.3cm}\mbox{and}
\hspace*{0.3cm}
\rho=\frac{\alpha(1-p'')}{\alpha+(1-\alpha)p''}
\end{equation}
and 
\begin{equation}
\alpha=\frac{\rho p}{1-p+\rho p}.
\end{equation}
$X_{ij}$ are given as
\begin{equation}
X_{ij}=(1-\alpha)p''^{i}(1-p'')^{j}+\alpha \delta_{j,0}
\end{equation}
$p_{ij}$ and $\rho_{ij}$ are calculated easily as
\begin{equation}
p_{i0}=\frac{\alpha+(1-\alpha)p''^{i+1}}{\alpha+(1-\alpha)p''^{i}}
\hspace*{0.3cm} \mbox{and}\hspace*{0.3cm}p_{ij}=p''
\hspace*{0.3cm}\mbox{for} 
\hspace*{0.3cm}j\neq 0 
\end{equation}
and 
\begin{equation}
\rho_{i0}=\frac{\alpha(1-p'')}{\alpha+(1-\alpha)p''^{i+1}}
\hspace*{0.3cm} \mbox{and}\hspace*{0.3cm}\rho_{ij}=0
\hspace*{0.3cm}\mbox{for} 
\hspace*{0.3cm}j\neq 0. 
\end{equation}

\end{itemize}

\subsection{Correlation Structures of the Solutions}

\begin{figure}[htbp]
\begin{center}
\includegraphics[width=12cm]{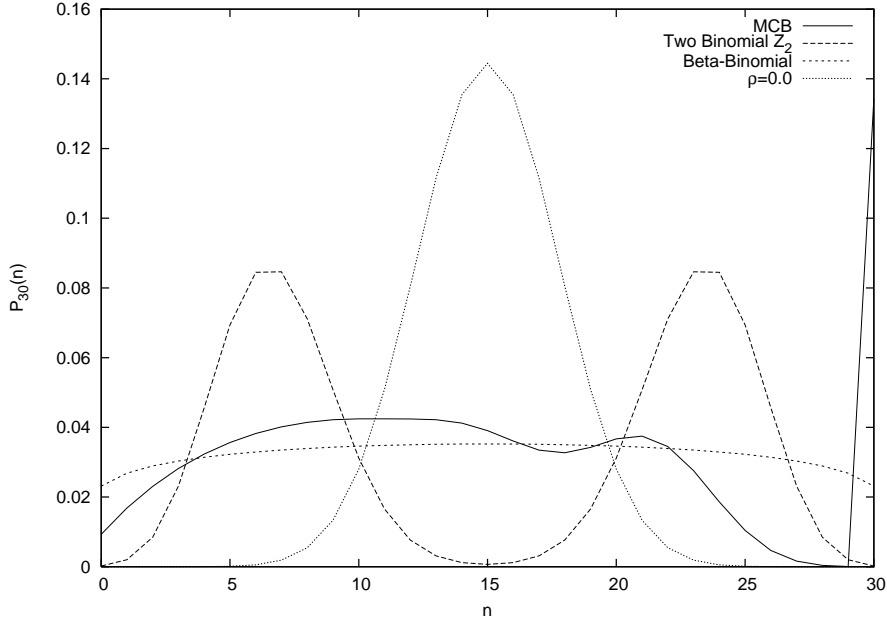} 
\caption{
Probability distribution $P_{30}(n)$ 
for $p=0.5,\rho=0.3$ and $N=30$.
We show 3 distributions, MCB (solid line), 
beta-binomial (dotted line) and Two-binomial (thin dotted line). We also
 show a  binomial distribution 
($\rho=0.0$) for comparison. 
}
\label{P_30_05_03}
\end{center}
\end{figure}

In this subsection, we  study the relations between probability
distributions and correlation structure.  
Figure \ref{P_30_05_03} shows the probability distribution 
profiles for three correlated models, MCB, 
BBD and Two-Binomial models. 
We set $p=0.5$, $\rho=0.3$ and $N=30$.
We also shows the pure binomial distribution for comparison.
The former three curves have the same $p$ and $\rho$, however their profiles
are drastically different.
 Two binomial model with $Z_{2}$ symmetry has two peaks and their
overlaps decreases as $N$ increases. At the thermodynamic limit 
$N \to \infty$, the overlap disappears and the system loses its 
ergodicity. The long-range Ising models shows spontaneous symmetry (SSB)
breaking of the $Z_{2}$ symmetry . 
On the other hand, the BBD's profile is broad and
 even if we set $N\to \infty$, we obtain the beta distribution 
and the shape is almost unchanged. That is, the BBD system does not show
SSB and it maintain its $Z_{2}$ (particle-hole) symmetry at $p=0.5$.

\begin{figure}[htbp]
\begin{center}
\includegraphics[width=8cm]{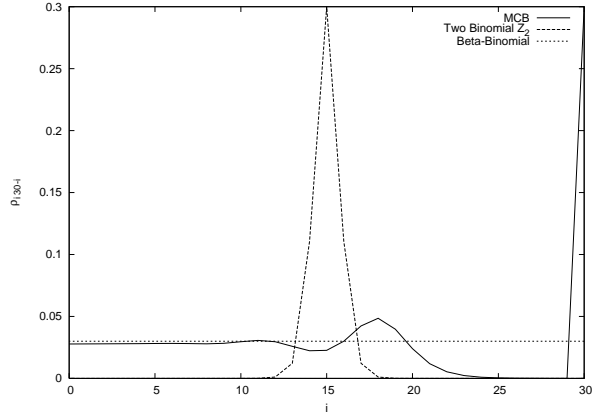} 
\caption{
Correlation $\rho_{i 30-i}$ for MCB (solid line), 
BBD (thin dotted line) and Two-Binomial (dotted line)
models. We set $\rho=0.3$ and $p=0.5$ as in the previous figure.
}
\label{Rho_30_05_03}
\end{center}
\end{figure}

The profile of MCB model is peculiar.  It is not symmetric and shows
 singular rippling. The origin for the ripping can be understood from the
 inspection of its correlation structure. 
Figure \ref{Rho_30_05_03}
 shows the correlation structures for the above three models. 
The
 parameters are equal and we show $\rho_{i 30-i}$.  In contrast 
to the BBD's correlation, which is constant with $i+j$ fixed, the 
correlations for MCB has sharp peak at $i=30$ and show strong 
rippling structure.  The curve is not symmetric and the distortion
is reflected in the shape of its probability distribution.
On the other hand, the correlation curve for Two-binomial distribution
has a strong peak at $i=\frac{N}{2}$ and is it much different from 
the BBD's correlation curve. This strong peak and rapid decay may be  
reflected in the decomposition of the probability distribution.
However, we have not yet understood the relation well.

\begin{figure}[htbp]
\begin{center}
\includegraphics[width=12cm]{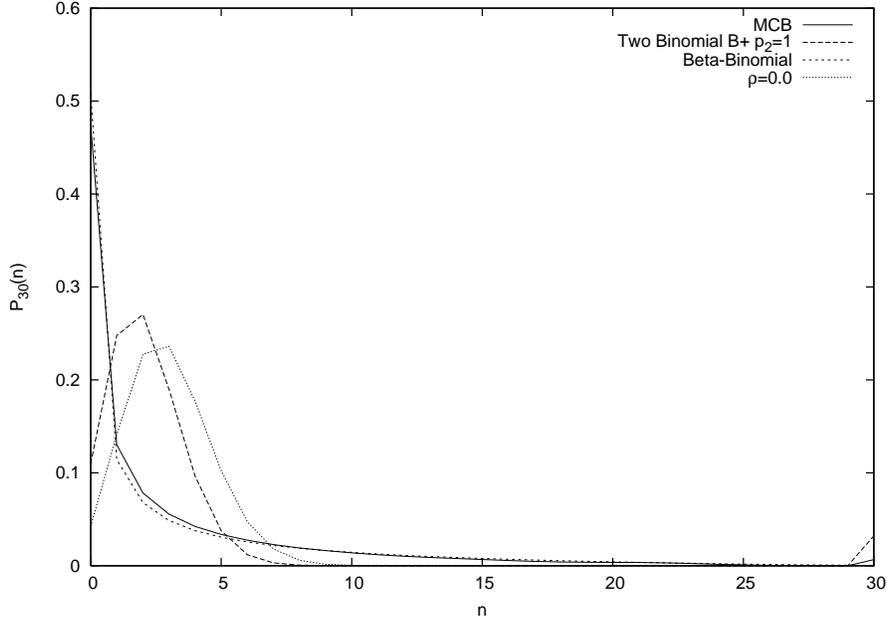} 
\caption{
Probability distribution $P_{30}(n)$ 
for $p=0.1,\rho=0.3$ and $N=30$.
We show 3 distributions, MCB (solid line), 
beta-binomial (dotted line) and B+ (thin dotted line). We also
 show a  binomial distribution 
($\rho=0.0$) for comparison. }
\label{P_30_01_03}
\end{center}
\end{figure}

\begin{figure}[htbp]
\begin{center}
\includegraphics[width=8cm]{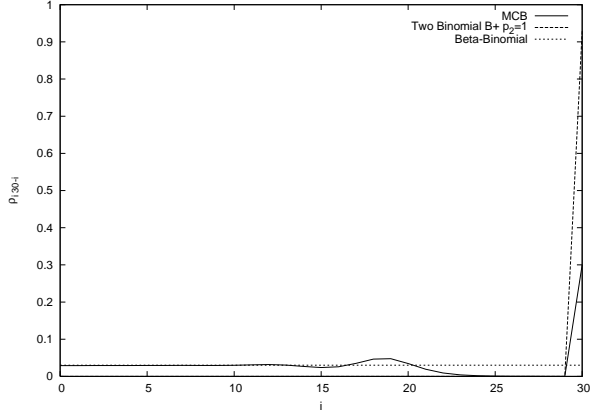} 
\caption{
Correlation $\rho_{i 30-i}$ for MCB (solid line), 
BBD (thin dotted line) and B+ (dotted line)
models. We set $\rho=0.3$ and $p=0.1$.
}
\label{Rho_30_01_03}
\end{center}
\end{figure}

Figure \ref{P_30_01_03} shows the probability distribution
for MCB, BBD and B+ models.
We set $p=0.1$, $\rho=0.3$ and $N=30$.
We also shows the pure binomial distribution for comparison.
MCB and BBD have almost the same bulk shape, however MCB
has a small peak at $n=30$. B+ has more strong peak at $n=30$ and its
bulk shape can be obtained by a small left shift of the pure binomial 
distribution $p=0.1$. These profile differences are reflected in their
correlation structures. See Figure \ref{Rho_30_01_03}. It 
 shows the correlation structures for the above three models. 
The parameters are equal as in the previous figure.  Contrary to the
constant BBD structure, MCB and B+ models have a peak at $i=30$.
MCB has a small and B+ has a tall peak and the difference is reflected
int the size of their tail peak of the probability distributions.

\section{Concluding Remarks and Future Problems}
\label{Conclusion}

In this paper, we show a general method to construct correlated
binomial models. We also estimate their moments.
Our method includes Witt's model and the BBD. 
In addition, with the consistent equations on $p_{ij}$ and 
$\rho_{ij}$, it is possible to prepare correlated binomial 
distributions with any choice for $\rho_{i0}$ or $p_{i0}$.
Of course, the resulting distribution function should be non-negative
, 'any' should be taken with some care. In addition, from the
joint probabilities $X_{ij}$, it is possible to estimate $p_{ij}$ and 
$\rho_{ij}$. We can see the detailed structure of the system with any
distribution function. In the work \cite{Bakkaloglu}, the conditional 
strange failure probabilities $p_{i0}$ were studied. 
Some recursive relations on $p_{i0}$ 
were  proposed and the resulting conditional probabilities
$p_{i0}$ were compared with real data on server networks.  
We note that $p_{i0}$ can be freely changed and it may be  possible
to make a good fitting with data. However, if the correlation structure
 $\rho_{ij}$ becomes too complex and  it shows oscillation, 
such a modeling may be over-fitting.

At last, we make comments about future problems. 
The first one is to seek another interesting
solution to eq.(\ref{consist_p}),
eq.(\ref{consist_q}) and eq.(\ref{consist3}) about $\rho_{ij}$ and
$p_{ij}$. In this paper, we have assumed strong symmetry
in $\rho_{ij}$ in the derivation of the BBD. For any value of $p$, 
we have assumed $Z_{2}$ symmetry
$\rho_{ij}=\rho_{ji}$. Furthermore, we have assumed stronger constraint that
$\rho_{ij}$ depends on $i,j$ only through the combination $i+j$.
The consistent relation is then solved easily and we get the
BBD. However, we think that 
the correlated binomial distribution space is rich and 
there may exist other interesting solutions. 
We discuss some simple solutions which are superpositions of 
two binomial distribution. They try to mimic the long-range Ising model
in the large $N$ limit and $\rho>0$ \cite{Kitsukawa}. 
A simple seamless solution for the consistent relations
which correspond to the long-range Ising model may exist.
Taking the continuous limit of the consistent relations and 
studying their solution is also an interesting problem.
The solution space may become narrow, however differential 
equations are more tractable than the recursion relations.
There should exist the beta distribution and 
the superposition of delta-functions, which are the continuous limits 
of the simple solutions presented here. 
    
The second problem is the generalization of the present method.
In this paper, we have assumed that the Bernoulli random variables
are all exchangeable. If one consider to apply the correlated binomial
model to the real world, such an idealization should be relaxed.
One possibility is the inhomogeneity in $p$ and the other is the
inhomogeneity in $\rho$. The first step is to add one other Bernoulli
random variable $Y$ to $N$ exchangeable variable system.
This $N+1$ system case has been treated in \cite{Mori}, it seems much
difficult to introduce the self-consistent equations in the present 
context. However, such a generalization may lead us to find new 
probability distribution functions, we believe that 
it deserves for extensive studies.

\bibliography{apssamp}

\begin{thebibliography}{99}


\bibitem{Griffiths} D.A.Griffiths, Biometrics {\bf 29} 637 (1973).

\bibitem{Williams} D.A.Williams, Biometrics {\bf 31} 949 (1975).

\bibitem{Kupper} L.L.Kupper and J.K.Haseman, Biometrics {\bf 34} 69 (1978). 


\bibitem{Bakkaloglu} M.Bakkaloglu et al, Technical Report CMU-CS-02-129,
	Carnegie Mellon University (2002).

\bibitem{Schonbucher}P.J.Sch\"{o}nbucher {\it Credit Derivatives
Pricing Models : Model, Pricing and Implementation}, 
U.S. John Wiley \& Sons (2003).

\bibitem{Frey} R. Frey and A. J. McNeil, Journal of Risk, {\bf 6} 59 (2003).

\bibitem{Witt} G.Witt, Moody's Correlated Binomial Default 
Distribution
 (Moody's Investors Service){\bf August 10} (2004). 

\bibitem{Mori}S.Mori, K.Kitsukawa and M.Hisakado,
Moody's Correlated Binomial Default Distributions for Inhomogeneous
	Portfolios, submitted to Quantitative Finance, arXiv:physics/0603036.

\bibitem{Kitsukawa}K.Kitsukawa, S.Mori and M.Hisakado, Evaluation of 
Tranche in Securitization  and Long-range Ising Model, to be published in
Physica {\bf A} (2006), arXiv:physics/0603040.


\bibitem{DeFinetti} De Finetti, {\it Theory of Probability}, Wiley (1974-5).

\bibitem{Goldenfeld} N. Goldenfeld, {\it Lectures on Phase Transitions and
	the Renormalization Group}, Addison-Wesley Publishing Company (1992).



\end{thebibliography}

\end{document}